\documentclass[12pt]{article}
\usepackage{graphicx}

\newcommand\pubdate{\today}

\def\Title#1{\begin{center} {\Large #1 } \end{center}}
\def\Author#1{\begin{center}{ \sc #1} \end{center}}
\def\Address#1{\begin{center}{ \it #1} \end{center}}

\newcommand\pubblock{\rightline{\begin{tabular}{l}  \\ 
         \pubdate  \end{tabular}}}
\newenvironment{Abstract}{\begin{quotation}  }{\end{quotation}}
\newenvironment{Presented}{\begin{quotation} \begin{center} 
             PRESENTED AT\end{center}\bigskip 
      \begin{center}\begin{large}}{\end{large}\end{center} \end{quotation}}

\begin{document}
\begin{titlepage}
~\\ \pubblock
\vfill
\Title{Overview of the ATLAS ALFA detectors, performance and physics analysis}
\vfill
\Author{P. J. Bussey}
\Address{SUPA-School of Physics and Astronomy, University of Glasgow, Glasgow G12 8QQ, U.K.\\[5mm] On behalf of the ATLAS Forward Detector group.}
\vfill
\begin{Abstract}
An overview is presented of the ALFA forward detector system in the
ATLAS detector at the LHC, CERN.  Details of the construction are
given, with summaries of the resulting analysis.
\end{Abstract}
\vfill
\begin{Presented}
DIS2023: XXX International Workshop on Deep-Inelastic Scattering and
Related Subjects, \\
Michigan State University, USA, 27-31 March 2023 \\[2mm]
     \includegraphics[width=9cm]{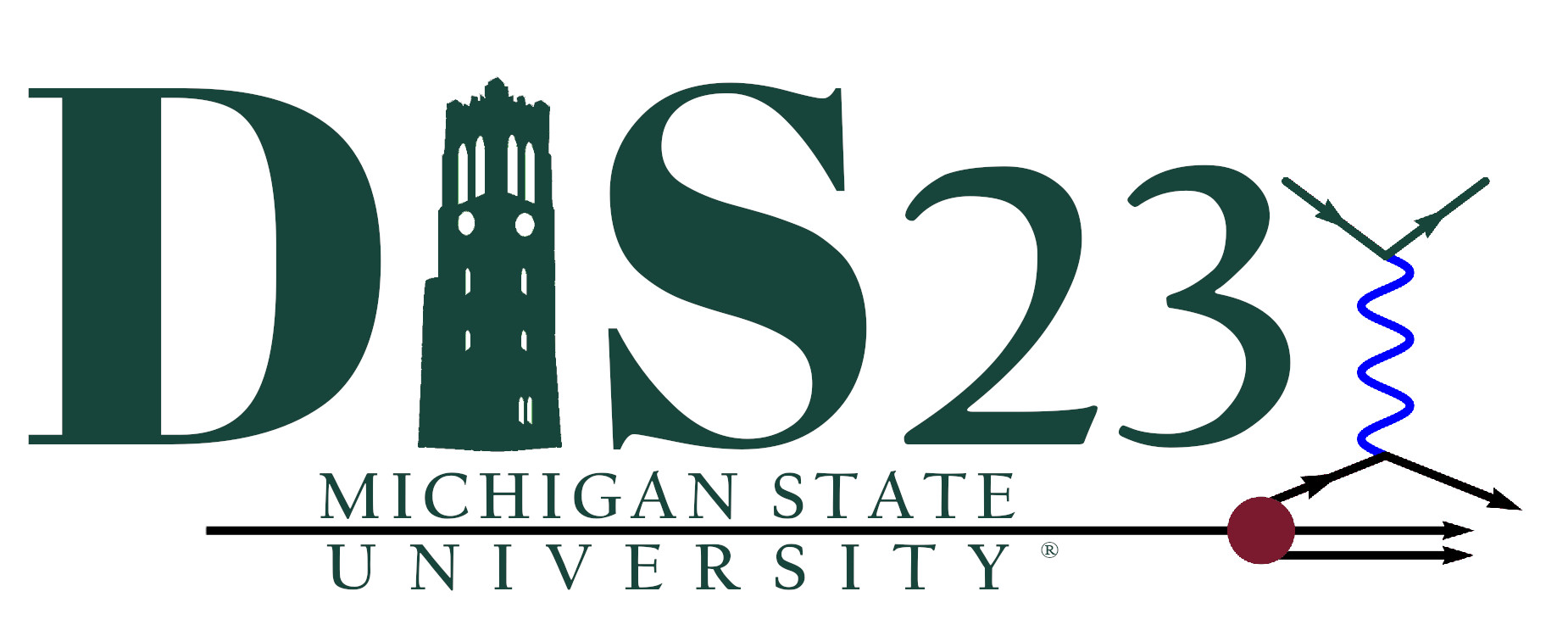}
\end{Presented}
\vfill
\end{titlepage}

The ATLAS experiment is a multi-purpose particle detection system
located at the interaction point IP1 in the LHC at CERN.  It comprises
a central detector, which covers a large range of particle
pseudorapidities centred on zero, and two forward detector systems,
which are able to measure protons that emerge at vertical angles close
to the beamline.  Figure 1 shows the layout of the ALFA stations
around 240~m from the interaction point~\cite{a1}; they are inserted
vertically and a second Roman Pot system (AFP), inserted horizontally,
is also in place and measures the deflection and energy loss of
emerging protons that are horizontally deflected by the dipole magnets
in the beamline~\cite{a2}.
\begin{figure}[b!]
  \includegraphics[width=0.9\textwidth]{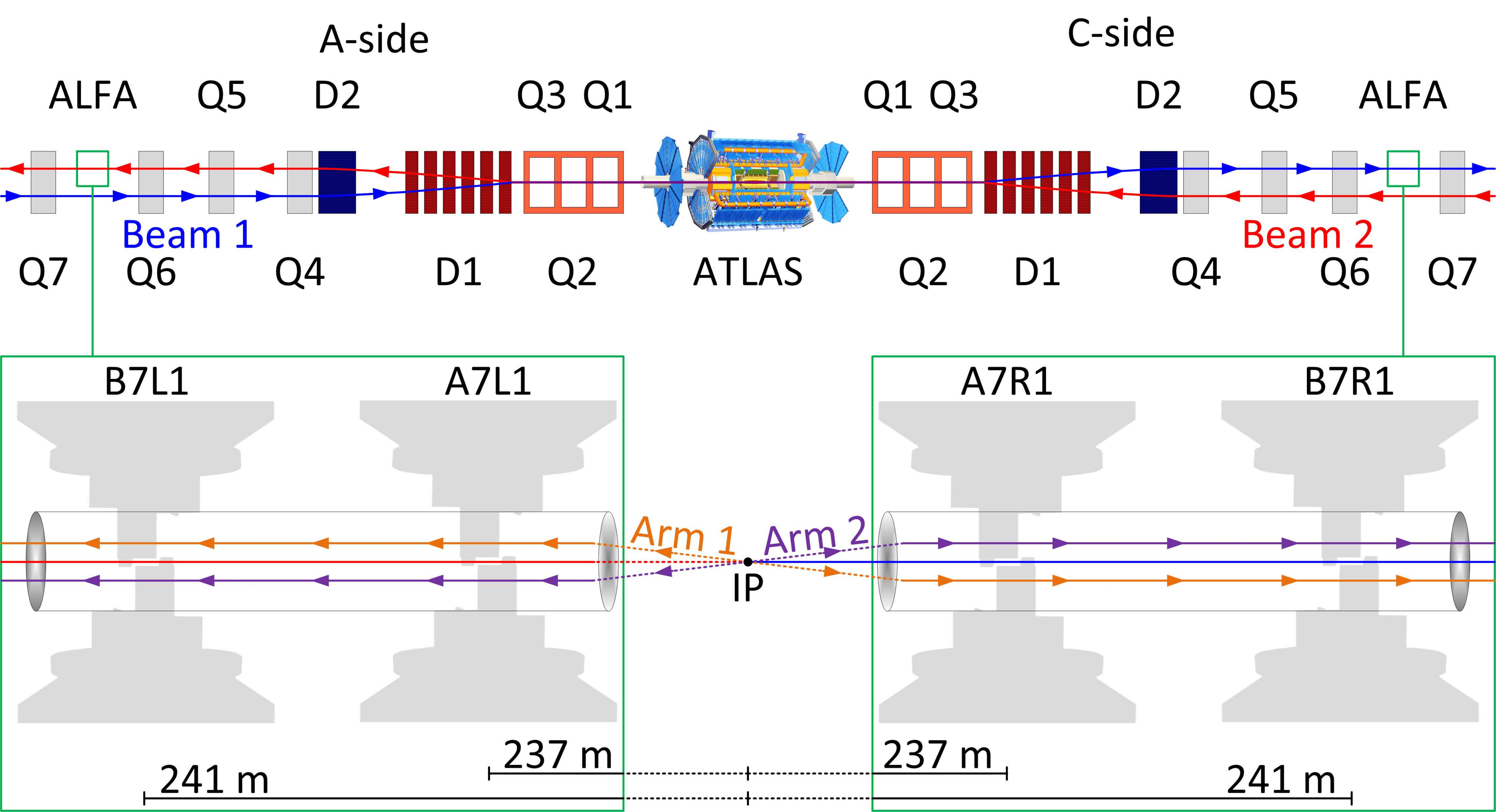}
\caption{Schematic view of the ATLAS detector and the positions of the
  ALFA stations relative to the beamline elements.  The upper
  diagram is a plan view; the lower diagram is a side projection
  indicating the paths of typical emerging vertically scattered proton pairs.  }
 \end{figure}

The ALFA system was originally proposed to measure the LHC
luminosity at the ATLAS interaction point (IP).  Since there are
superior ways of measuring luminosity, the ALFA system has been exploited
primarily to measure elastically scattered proton pairs, together with
doubly-diffractive events with two very forward protons accompanied by
an exclusive centrally produced hadronic system.  In order to measure
protons scattered through extremely small angles, special LHC beam
optics settings are required in dedicated runs with high values of
$\beta^\star$ at the IP, so as to give parallel-to-point focussing at
the ALFA stations.  These runs are quite short, with low luminosity,
and are not compatible with the normal high-luminosity running of the
LHC.

Further details of the construction of the ALFA detectors are shown in
Figures 2 and 3~\cite{a1,r5}. In each station, detectors are mounted
in two Roman Pot systems that are inserted close to the beamline from
above and below.  Protons are detected by means of arrays of
scintillating fibres.  Three types of fibre arrays are employed.  Two
of these arrays have the fibres oriented at $\pm45^\circ$ relative to
the vertical, in order to measure corresponding coordinates denoted as
$u$ and $v$.  The third type of detector consists of detector planes
with horizontal fibres that are capable of measuring a subset of
protons that also pass through both the upper and lower detector
systems.  These measurements are used to provide information on the
relative positions of the upper and lower modules while running.  One
overlap detector is included in each of the upper and lower ALFA
modules.  A relative alignment accuracy of 5-10 $\mu$m was achieved.
After the relative alignment was optimised, a global vertical position
uncertainty remains of $\pm22\;\mu$m.

\begin{figure}[t]
\includegraphics[width=0.55\textwidth]{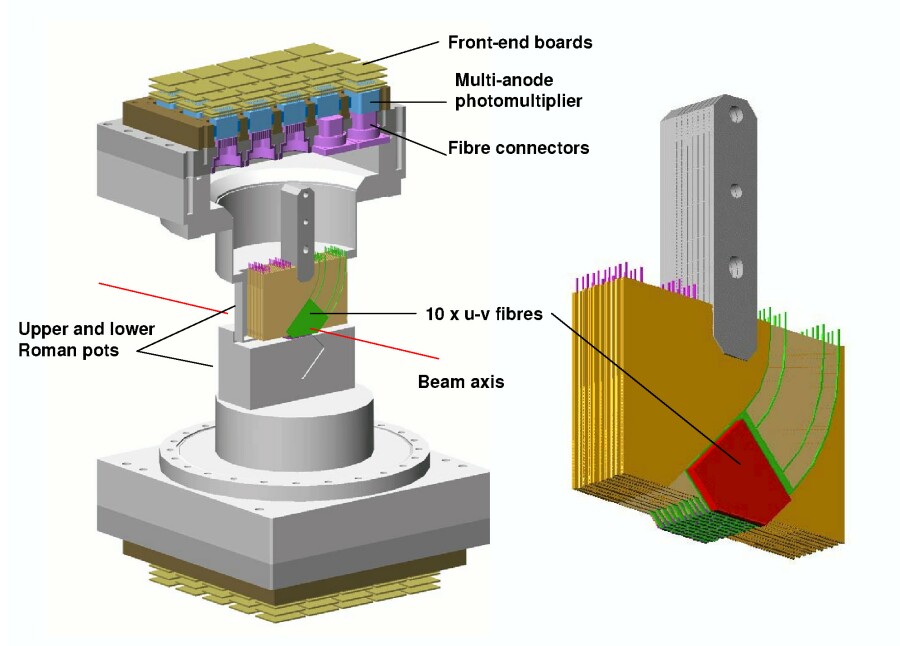}
\includegraphics[width=0.43\textwidth]{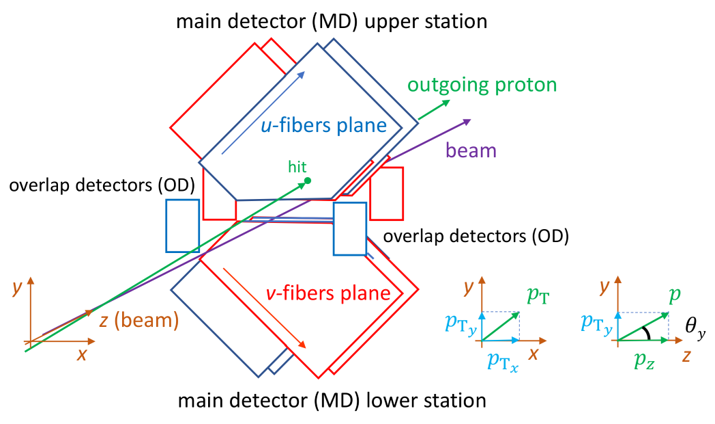}
\caption{(Left) Configuration of the detectors and readout 
in one ALFA Roman Pot; (right) layout of the $u$ and $v$ measuring detectors
in the two Roman pots of one station, together with the overlap detectors.
}
 \end{figure} 

\begin{figure}
\includegraphics[width=0.4\textwidth]{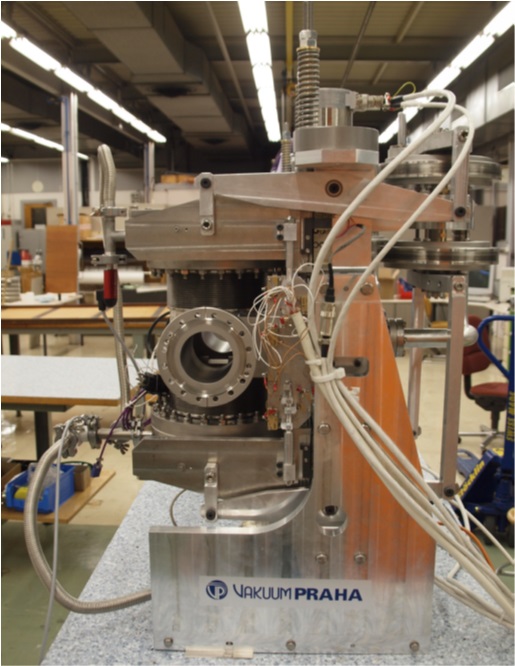}
\includegraphics[width=0.59\textwidth]{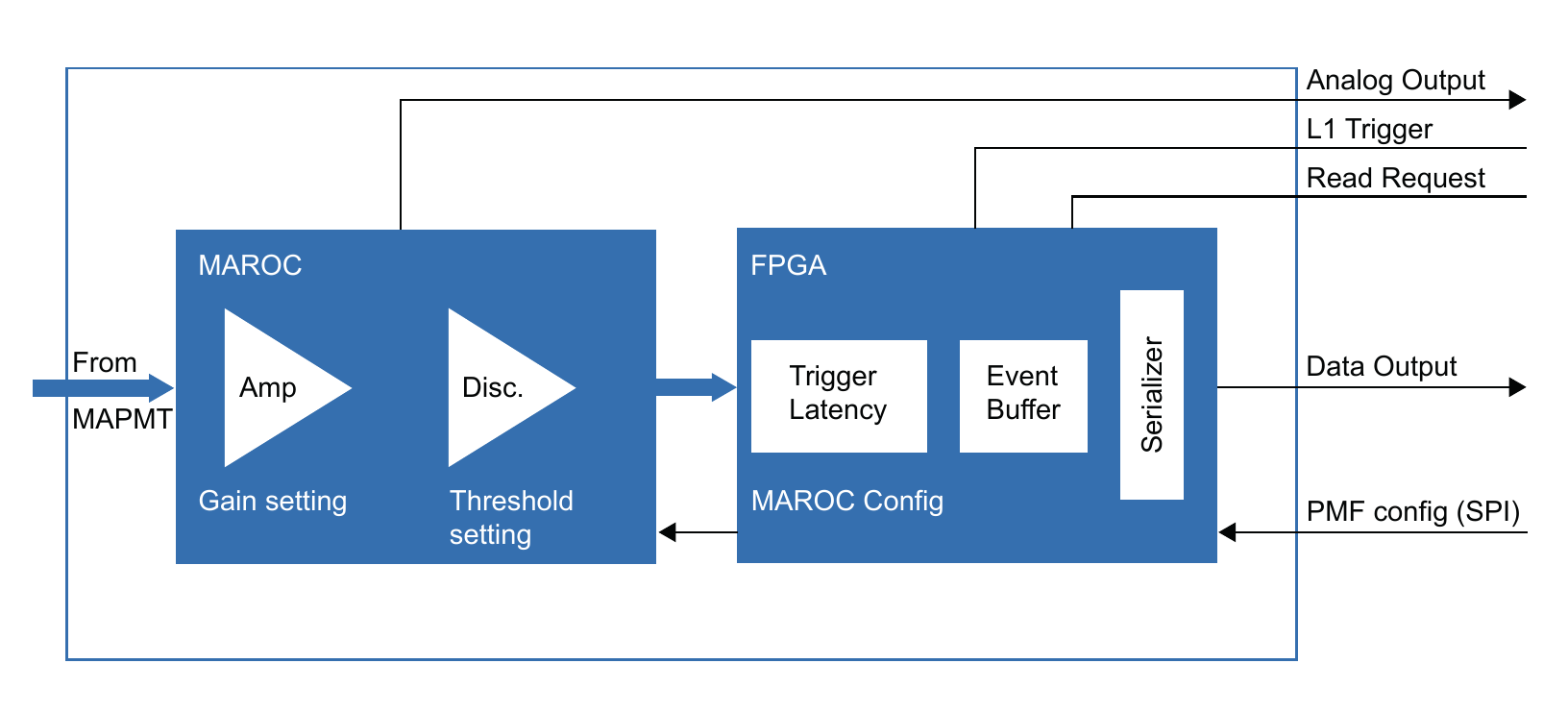}
\caption{ (Left) mounting of an ALFA 
station in a dedicated section of the LHC beam pipe.
(Right) schematic diagram of the trigger and readout system for ALFA. 
}
\end{figure}

The scintillating fibres have a square cross-section of side 0.5 mm,
and are read out by means of Multi-Anode Photomultiplier Tubes.
Trigger counters, which are 3 mm thick plastic scintillator tiles, are
located between the overlap detectors and the main detectors in each
of the upper and lower modules.  The scintillating fibres have limited
radiation tolerance, which itself restricts their operation to
low-luminosity running of the LHC.

Figure 3 (right)  shows schematically the design of the front-end electronics
of the system.  A first-level trigger signal is supplied from each
ALFA module.  In particular, so-called ``elastic'' proton-proton
combinations, as indicated by Arm1 or Arm2 in Figure 1 may be
triggered on, and also ``anti-elastic'' combinations which are defined
as signals from modules (A1 or A3) and (A5 or A7) , or (A2 or A4)
and (A6 or A8).  The anti-elastic combinations had a high rate and
were prescaled.

\begin{figure}
\includegraphics[width=0.45\textwidth]{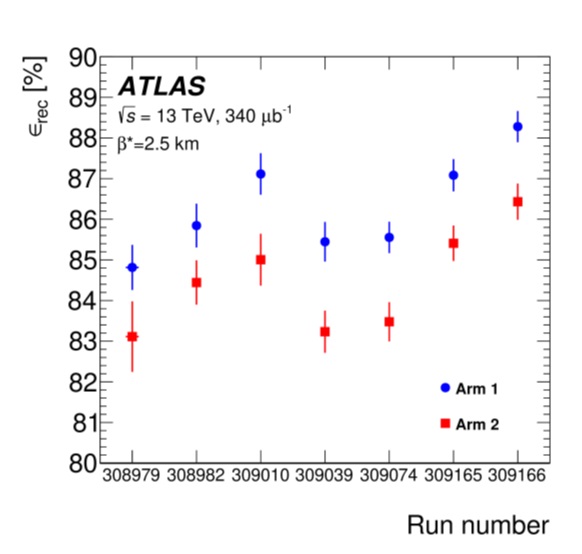}
\includegraphics[width=0.45\textwidth]{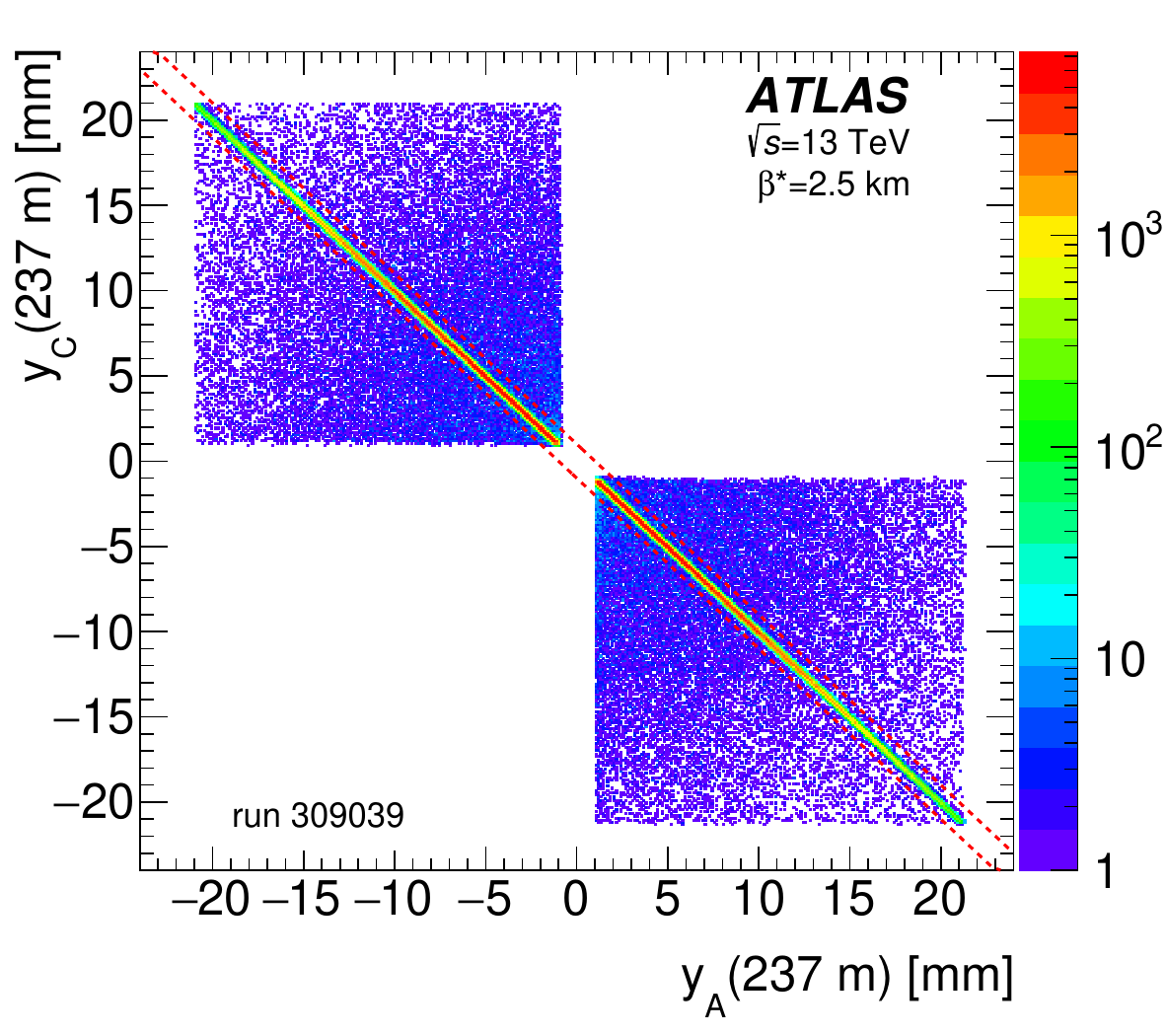}
\caption{(Left) event reconstruction efficiency of the ALFA system for
  proton detection for a series of runs;
  (right) reconstructed proton trajectory pairs in the A and C sides
  of the ATLAS system.  }
 \end{figure}

The reconstruction efficiencies of the system were measured by a
tag-and-probe method: well measured protons on one side of ATLAS were
used as tags to measure the efficiency for measuring a proton on the
other side.  The results are illustrated in Figure 4~\cite{r5}, 
which also illustrates the well correlated nature of proton pairs in
mainly elastic events detected in the two sections of the arms.

A series of precise measurements of elastic proton-proton scattering
have been published~\cite{r1,r2,r3}. A first measurement of exclusive
pion pairs in double-diffractive proton scattering has also been
published~\cite{r5}. While of limited statistical accuracy, the
observed signal in the elastic channel has very little background
after a set of cleaning cuts is made on the data (Figure 5).

\begin{figure}
\begin{center}
~\\[-2.5mm]
\includegraphics[width=0.5\textwidth]{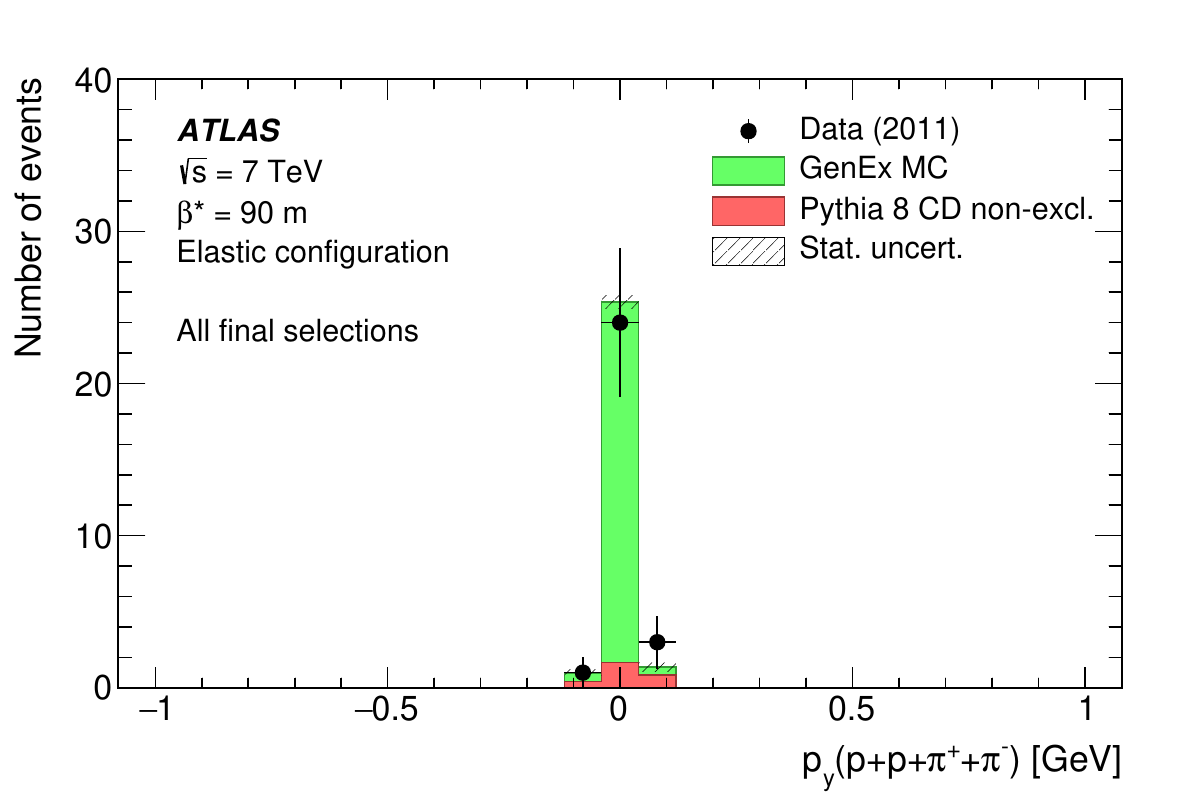}
\caption{The total vertical momentum of the outgoing $pp\pi^+\pi^-$ system
with two ALFA proton tags and an exclusive central pion pair.
}
\end{center}
~\\[-11mm]
 \end{figure}

Further data for exclusive process are in the process of being
analysed.  The final ALFA operation is scheduled in the LHC Run 3.
This will enable more precision on the elastic measurements and higher
precision on exclusive processes, to enable detailed analysis of
Pomeron-Pomeron scattering final states to be performed.

\noindent
Copyright CERN for the benefit of the ATLAS Collaboration. CC-BY-4.0 license.

\end{document}